\documentclass[11pt]{article}
\usepackage{amsmath}
\usepackage{amsfonts}
\usepackage{amssymb}
\usepackage{graphicx}
\usepackage{bm}
\usepackage{color}
\usepackage{amscd}

\def\bea{\begin{eqnarray}}
\def\eea{\end{eqnarray}}
\begin{document}
\begin{center}
\LARGE {\bf Scalar perturbation in warm tachyon inflation in LQC in light of Plank and BICEP2}
\end{center}
\begin{center}
{M. R. Setare $^{a}$\footnote{E-mail: rezakord@ipm.ir
}\hspace{1mm} ,
V. Kamali $^{b}$\footnote{E-mail: vkamali1362@gmail.com, Vkamali@basu.ac.ir}\hspace{1.5mm} \\
 $^a$ {\small {\em Department of Science, Campus of Bijar, University of Kurdistan.\\
  Bijar , Iran.}}\hspace{1.5mm}\\
$^b$ {\small {\em  Department of Physics, Faculty of Science,\\
Bu-Ali Sina University, Hamedan, 65178, Iran}}}\\

\end{center}

%\vskip 3cm

\begin{center}
{\bf{Abstract}}\\
We study warm-tachyon inflationary universe model in the context
of the effective field theory of loop quantum cosmology. In
slow-roll approximation the primordial perturbation spectrums
for this model are calculated. We also obtain the general
expressions of the tensor-to-scalar ratio, scalar spectral index.
We develop this model by using exponential
potential, the characteristics of this model is calculated in
great details. The parameters of the model are restricted by
recent observational data from Planck,  WMAP9 and BICEP2.
 \end{center}

\newpage

\section{Introduction}
 Big Bang model have many long-standing problems (horizon,
flatness,...). These problems are solved in the framework of the
inflationary universe models \cite{1-i}. Scalar field as a source
of inflation provides the causal interpretation of the origin of
the distribution of large scale structure and observed anisotropy
of cosmological microwave background (CMB) \cite{6}. In standard
models for inflationary universe, the inflation period is divided
into two regimes, slow-roll and reheating epochs. In slow-roll
period kinetic energy remains small compared to the potential
terms. In this period, all interactions between scalar fields
(inflatons) and  other fields are neglected and the universe
inflates. Subsequently, in reheating period, the kinetic energy
is comparable to  the potential energy and inflaton starts an
oscillation around the minimum of the potential losing their
energy to other fields present in the theory. So, the reheating
is the end period of inflation.\\ In warm inflationary models
radiation production occurs during inflationary period and
reheating is avoided \cite{3}. Thermal fluctuations may be
obtained during warm inflation. These fluctuations could play a
dominant role to produce initial fluctuations which are necessary
for Large-Scale Structure (LSS) formation. So the density
fluctuation arises from thermal rather than quantum fluctuation
\cite{3-i}. Warm inflationary period ends when the universe stops
inflating. After this period the universe enters in radiation
phase smoothly \cite{3}. Finally, remaining inflatons or dominant
radiation fields created the matter components of the universe.\\
Friedmann-Robertson-Walker (FRW) cosmological models in the
context of string/M-theory have related to brane-antibrane
configurations \cite{4-i}. Tachyon fields associated with
unstable D-branes may be responsible for inflation in early time
\cite{5-i}. The tachyon inflation is a k-inflation model
\cite{n-1} for scalar field $\phi$ with a positive potential
$V(\phi)$. Tachyon potentials have two special properties,
firstly a maximum of these potential is obtained where
$\phi\rightarrow 0$ and second property is the minimum of these
potentials is obtained where $\phi\rightarrow \infty$. If the
tachyon field start to roll down the potential, then universe
dominated by a new form of matter will smoothly evolve from
inflationary universe to an era which is dominated by a
non-relativistic fluid \cite{1}. So, we could explain the phase
of acceleration expansion (inflation) in term of tachyon field.\\
The warm tachyon inflationary model  have been studied in
Ref.\cite{1-m}. In the present work we will study warm-tachyon
inspired inflation in the context of the effective theory of loop
quantum gravity (LQG). Techniques of LQG which is a resulting
non-perturbative background independent approach to quantizing
gravity \cite{8-i}, could be applied in homogeneous and isotropic
space-time which is known as loop quantum cosmology (LQC).
Canonical quantization gravity in term of Ashtekar-Barbero
connection variables is studied in LQG. In LQG the phase space of
classical general relativity may be spanned by conjugate
variables $A_a^i$ (connection) and $E_i^a$ (triad) on a
3-manifold $\mathcal{M}$ which encode curvature and spatial
geometry respectively(labels $a$ and $i$ denote internal indices
of $SU(2)$ and space index respectively). Due to the isotropic
and homogeneous symmetries, in LQC model the phase space is
simplified. The phase space of this model is spanned by a single
connection $\textbf{c}$ and a single triad $\textbf{p}$. The
Poisson bracket for LQC variable is given by
\begin{equation}\label{}
\nonumber \{\textbf{c},\textbf{p}\}=\frac{8\pi\gamma}{3m_p^2}
\end{equation}

where $\gamma$ is the dimensionless Barbere-Immirzi parameter.
For spatially flat (FRW) universe the LQC variables \textbf{c}
and \textbf{p} have these relations with the metric components
\begin{equation}\label{}
\nonumber \textbf{c}=\gamma\dot{a}~~~~~~~\textbf{p}=a^2
\end{equation}

Classical Hamiltonian constraint in term of connection and triad
variables is given by
\begin{equation}\label{}
\nonumber
\mathcal{H}_{cl}=-\frac{3\sqrt{\textbf{p}}}{\gamma^2}+\textbf{c}^2+\mathcal{H}
\end{equation}

where $\mathcal{H}_m$ is the matter Hamiltonian. In Hamiltonian
formalism, the dynamical equations (modified Friedmann equation)
may be determined by the above Hamiltonian constraint. The
effective classical Hamiltonian constraint in terms of
kinematical length of the edge of square loop $\overline{\mu}$ is
given by \cite{9-i}
\begin{equation}\label{}
\nonumber
\mathcal{H}=-\frac{3}{\gamma\overline{\mu}^2}a\sin^2(\mathcal{\mu}\textbf{c})+\mathcal{H}_m\\
\end{equation}

Using the Hamilton equation of motion
\begin{equation}\label{}
\nonumber
\dot{\textbf{p}}=\{\textbf{p},\mathcal{H}_{eff}\}=-\frac{\gamma}{3}\frac{\partial
\mathcal{H}_{eff}}{\partial \textbf{c}}
\end{equation}

and the vanishing of the Hamiltonian constraint
($\mathcal{H}_{eff}\approx 0$) \cite{9-i}, these two important
relations are obtained
\begin{eqnarray}\label{}
\nonumber
\dot{a}=\frac{1}{\gamma\overline{\mu}}\sin(\overline{\mu} \textbf{c})\cos(\overline{\mu} \textbf{c})\\
\nonumber \sin^2(\overline{\mu} \textbf{c})=\frac{8\pi}{3m_p^2
a}\mathcal{H}_m
\end{eqnarray}

Therefore, from above equation the modified Friedmann equation
becomes
\begin{equation}\label{g}
(\frac{\dot{a}}{a})^2=\frac{8\pi}{3m_p^2}\rho[1-\frac{\rho}{\rho_c}]~~~~~~~~\rho_c=\frac{4\sqrt{3}}{\gamma^3}
\end{equation}

In this paper we will study warm-tachyon inflationary model in
the context of LQC by using the above modified Friedmann
equation. The paper organized as: In the next section we will
describe warm-tachyon inflationary universe model in the
framework of LQC. In section (3) we consider the perturbations
for our model and obtain scalar and tensor perturbation
spectrums. In section (4) we study our model  using the
exponential potential in high dissipative regime. Finally in
section (5) we close by some concluding  remarks.

\section{The model}
In the present work we will study warm-tachyon inspired inflation
in the context of effective field theory of LQC where the
modified Friedmann equation has the following form
\begin{equation}\label{1}
H^2=\frac{1}{3}[\rho_{\phi}+\rho_{\gamma}][1-\frac{\rho_{\phi}+\rho_{\gamma}}{\rho_{c}}]
\end{equation}

where $H=\frac{\dot{a}}{a}$ is the Hubble factor, $a$ is the
scale factor and we choose $c=\hbar=8\pi G=\frac{8\pi}{m_p^2}=1$
($m_p$ is Planck mass.). Energy-momentum tensor of tachyonic
inflation model in a spatially flat Friedmann Robertson Walker
(FRW) is recognized by
$T_{\mu}^{\nu}=diag(-\rho_{\phi},P_{\phi},P_{\phi},P_{\phi})$
where the pressure and energy density of tachyon field are
defined by \cite{1}
\begin{eqnarray}\label{2}
P_{\phi}=-V(\phi)\sqrt{1-\dot{\phi}^2}
\end{eqnarray}

and
\begin{eqnarray}\label{3}
\rho_{\phi}=\frac{V(\phi)}{\sqrt{1-\dot{\phi}^2}}~~~~~~~
\end{eqnarray}

respectively, where $V(\phi)$ is the effective scalar potential
associated with tachyon field $\phi$. Important characteristics
of this potential are $\frac{dV}{d\phi}<0$ and $V(\phi\rightarrow
0)\rightarrow V_{max}$ \cite{2}. The dynamic of warm tachyon
inflation in spatially flat FRW model in the context of effective
theory LQC is described by these equations.
\begin{eqnarray}\label{4}
H^2=\frac{1}{3}[\frac{V(\phi)}{\sqrt{1-\dot{\phi}^2}}+\rho_{\gamma}][1-\frac{1}{\rho_c}(\frac{V(\phi)}{\sqrt{1-\dot{\phi}^2}})]
\end{eqnarray}

\begin{equation}\label{5}
\dot{\rho}_{\phi}+3H(P_{\phi}+\rho_{\phi})=-\Gamma\dot{\phi}^2\Rightarrow
\frac{\ddot{\phi}}{1-\dot{\phi}^2}+3H\dot{\phi}+\frac{V'}{V}=-\frac{\Gamma}{V}\sqrt{1-\dot{\phi}^2}\dot{\phi}
\end{equation}

and
\begin{equation}\label{6}
\dot{\rho}_{\gamma}+4H\rho_{\gamma}=\Gamma\dot{\phi}^2
\end{equation}

where $\rho_{\gamma}$ is the energy density of the radiation and
$\Gamma$ is the dissipative coefficient with the dimension
$m_{p}^{5}$. In the above equations dots "." mean derivative with
respect to cosmic time and prime  denotes derivative with respect
to scalar field $\phi$. During inflation epoch the energy density
(\ref{3}) is the order of potential $\rho_{\phi}\sim V$ and
dominates over the radiation energy $\rho_{\phi}>\rho_{\gamma}$.
Using slow-roll approximation when $\dot{\phi}\ll 1$ and
$\ddot{\phi}\ll(3H+\frac{\Gamma}{V})$ \cite{3} and when inflation
radiation production is quasi-stable ($\dot{\rho}_{\gamma}\ll
4H\Gamma$, $\dot{\rho}_{\gamma}\ll\Gamma\dot{\phi}^2$) the
dynamic equations (\ref{4}) and (\ref{5})   are reduced to
\begin{equation}\label{7}
H^2=\frac{1}{3}V(1-\frac{V}{\rho_c})
\end{equation}

\begin{equation}\label{8}
3H(1+r)\dot{\phi}=-\frac{V'}{V}
\end{equation}

where $r=\frac{\Gamma}{3HV}$. From above equations and
Eq.(\ref{6}), $\rho_{\gamma}$ could be written as
\begin{equation}\label{9}
\rho_{\gamma}=\frac{\Gamma\dot{\phi}^2}{4H}=\frac{r}{4(1+r)^2(1-\frac{V}{\rho_c})}(\frac{V'}{V})^2=\sigma
T_r^4
\end{equation}

where  $T_r$ is the temperature of thermal bath and $\sigma$ is
Stefan-Boltzmann constant. We introduce the slow-roll parameters
for our model as
\begin{eqnarray}\label{10}
\epsilon=-\frac{\dot{H}}{H^2}\simeq\frac{V'^2}{2(1+r)V^3}\frac{1-\frac{2V}{\rho_c}}{1-\frac{V}{\rho_c}}
\end{eqnarray}

and
\begin{eqnarray}\label{11}
\eta=-\frac{\ddot{H}}{H\dot{H}}\simeq\frac{2V'}{V^2(1+r)[1-\frac{V}{\rho_c}]}~~~~~~~~~~~~~~~~~~~~\\
\nonumber \times
[\frac{V''}{V'}-\frac{V'}{V}-\frac{r'}{2(1+r)}-\frac{V'}{\rho_c-2V}+\frac{V'}{2\rho_c-2V}]
\end{eqnarray}

A relation between two energy densities $\rho_{\phi}$ and
$\rho_{\gamma}$ is obtained from Eqs. (\ref{9}) and (\ref{10})
\begin{equation}\label{12}
\rho_{\gamma}=\frac{r}{2(1+r)}\frac{[1-\frac{\rho_{\phi}}{\rho_c}]}{[1-\frac{2\rho_{\phi}}{\rho_c}]}\rho_{\phi}\epsilon\simeq\frac{r}{2(1+r)}\frac{[1-\frac{V}{\rho_c}]}{[1-\frac{2V}{\rho_c}]}V\epsilon
\end{equation}

The condition of inflation epoch $\ddot{a}>1$ could be obtained
by inequality $\epsilon<1$. Therefore from above equation,
warm-tachyon inflation in the context of effective theory LQC
could take place when
\begin{equation}\label{13}
\frac{2(1+r)}{r}\rho_{\gamma}<\frac{1-\frac{\rho_{\phi}}{\rho_c}}{1-\frac{2\rho_{\phi}}{\rho_c}}
\end{equation}

Inflation period ends when $\epsilon\simeq 1$ which implies
\begin{equation}\label{14}
[\frac{V'_f}{V_f}]^2\frac{1-\frac{2V}{\rho_c}}{1-\frac{V}{\rho_c}}\frac{1}{V_f}\simeq
2(1+r_f)
\end{equation}

where the subscript $f$ denotes the end of inflation. The number
of e-folds is given by
\begin{eqnarray}\label{15}
N=\int_{\phi_{*}}^{\phi_f}Hdt=\int_{\phi_{*}}^{\phi_f}\frac{H}{\dot{\phi}}d\phi=-\int_{\phi_{*}}^{\phi_f}\frac{V^2}{V'}(1+r)[1-\frac{V}{\rho_c}]d\phi
\end{eqnarray}

where the subscript $*$ denotes the epoch when the cosmological
scale exits the horizon.
\section{Perturbation}
 In quantum cosmology the interesting primary quantities are the curvature
and tensor perturbation spectrums which may be extracted from
two-point function of two quantum fields in the same time. In this section we will study the cosmological perturbations for
our model in high dissipative regime ($r\gg 1$) that lead to the  perturbation spectrums \cite{1-mo}.
 Sclar perturbations in the longitudinal gauge, may be described by the perturbed FRW metric
\begin{equation}\label{16}
ds^2=(1+2\Phi)dt^2-a^2(t)(1-2\Psi)\delta_{ij}dx^idx^j
\end{equation}

where $\Phi$ and $\Psi$ are gauge-invariant metric perturbation variables \cite{1-mo}.
The equation of motion is given by

 \begin{eqnarray}\label{}
\frac{\ddot{\delta\phi}}{1-\dot{\phi}^2}+[3H+\frac{\Gamma}{V}]\dot{\delta\phi}+[-a^{-2}\nabla^2+(\ln V)''+\dot{\phi}(\frac{\Gamma}{V})']\delta\phi\\
\nonumber
-[\frac{1}{1-\dot{\phi}^2}+3]\dot{\phi}\dot{\Phi}-[\dot{\phi}\frac{\Gamma}{V}-2(\ln V)']\Phi=0
\end{eqnarray}
We expand the small change of field $\delta\phi$ into Fourier components as
\begin{eqnarray}\label{}
\delta\phi(x)=\int \frac{d^3k}{(2\pi)^3}[e^{ikx}\delta\phi(k,t)\overrightarrow{a}_k+e^{-ikx}\delta\phi(k,t)\overrightarrow{a}_k^{\dag}]
\end{eqnarray}
where $a_{k}$ and $a_{k}^{\dag}$ denote the annihilation and creation operators respectively. These operators obey the simple commutation relations
\begin{eqnarray}\label{}
[a_k,a_{k'}^{\dag}]=(2\pi)^3\delta^3(k-k')~~~~~~~~~~~[a_k,a_{k'}]=0=[a^{\dag}_k,a^{\dag}_{k'}]
\end{eqnarray}

All perturbed quantities have a spatial sector of the form $e^{i\mathbf{kx}}$, where $k$ is the wave number. Perturbed Einstein field equations in momentum space have only the temporal parts
\begin{equation}\label{}
\nonumber
\Phi=\Psi
\end{equation}

\begin{equation}\label{17}
\dot{\Phi}+H\Phi=\frac{1}{2}[-\frac{4\rho_{\gamma}av}{3k}+\frac{V\dot{\phi}}{\sqrt{1-\dot{\phi}^2}}\delta\phi][1-\frac{2}{\rho_c}[\rho_{\gamma}+\frac{V}{\sqrt{1-\dot{\phi}^2}}]]
\end{equation}

\begin{eqnarray}\label{18}
\frac{\ddot{\delta\phi}}{1-\dot{\phi}^2}+[3H+\frac{\Gamma}{V}]\dot{\delta\phi}+[\frac{k^2}{a^2}+(\ln V)''+\dot{\phi}(\frac{\Gamma}{V})']\delta\phi\\
\nonumber
-[\frac{1}{1-\dot{\phi}^2}+3]\dot{\phi}\dot{\Phi}-[\dot{\phi}\frac{\Gamma}{V}-2(\ln V)']\Phi=0
\end{eqnarray}

\begin{eqnarray}\label{19}
(\dot{\delta\rho_{\gamma}})+4H\delta\rho_{\gamma}+\frac{4}{3}ka\rho_{\gamma}v-4\rho_{\gamma}\dot{\Phi}-\dot{\phi}^2\Gamma'\delta\phi-\Gamma\dot{\phi}^2[2(\dot{\delta\phi})-3\dot{\phi}\Phi]=0
\end{eqnarray}

and
\begin{equation}\label{20}
\dot{v}+4Hv+\frac{k}{a}[\Phi+\frac{\delta\rho_{\gamma}}{4\rho_{\gamma}}+\frac{3\Gamma\dot{\phi}}{4\rho_{\gamma}}\delta\phi]
\end{equation}
The above equations are obtained for Fourier components $e^{i\mathbf{kx}}$, where the subscript $k$ is omitted. $v$ in the above set of equations is given by the decomposition of the velocity field ($\delta u_j=-\frac{iak_J}{k}ve^{i\mathbf{kx}}, j=1,2,3$) \cite{1-mo}.

Warm inflation models could be considered as a hybrid-like inflationary model where inflaton field interacts with radiation field \cite{9-f}, \cite{8-f}. Entropy perturbation  relates to dissipation term \cite{10-f}. During slow-roll inflationary phase, for non-decreasing adiabatic modes on large scale limit $k\ll aH$, we assume that the perturbed quantities do not vary strongly. So we constrain above equation as: $H\Phi\gg\dot{\Phi}$, $(\ddot{\delta\phi})\ll(\Gamma+3H)(\dot{\delta\phi})$, $(\dot{\delta\rho_{\gamma}})\ll\delta\rho_{\gamma}$ and $\dot{v}\ll 4Hv$. In the slow-roll limit, and by using the above limitations, the set of perturbed equations reduce  to
\begin{equation}\label{21}
\Phi\simeq\frac{1}{H2}[-\frac{4\rho_{\gamma}av}{3k}+V\dot{\phi}\delta\phi][1-\frac{2V}{\rho_c}]
\end{equation}

\begin{equation}\label{22}
[3H+\frac{\Gamma}{V}]\dot{\delta\phi}+[(\ln V)''+\dot{\phi}(\frac{\Gamma}{V})']\delta\phi
\simeq[\dot{\phi}\frac{\Gamma}{V}-2(\ln V)']\Phi
\end{equation}

\begin{equation}\label{23}
\frac{\delta\rho_{\gamma}}{\rho_{\gamma}}\simeq\frac{\Gamma'}{\Gamma}\delta\phi-3\Phi
\end{equation}
and
\begin{eqnarray}\label{24}
v\simeq-\frac{k}{4aH}(\Phi+\frac{\delta\rho_{\gamma}}{4\rho_{\gamma}}+\frac{3\Gamma\dot{\phi}}{4\rho_{\gamma}}\delta\phi)
\end{eqnarray}
Using Eqs.(\ref{21}), (\ref{23}) and (\ref{24}) we determine the perturbation variable $\Phi$:
\begin{eqnarray}\label{25}
\Phi=\frac{V\dot{\phi}}{2H}[1+\frac{\Gamma}{4HV}+\frac{\Gamma'\dot{\phi}}{48H^2V}](1-\frac{2V}{\rho_c})\delta\phi
\end{eqnarray}

We can solve the above equations by taking inflaton $\phi$ as the independent variable in place of cosmic time $t$. Using Eq.(\ref{8}) we find
\begin{eqnarray}\label{26}
(3H+\frac{\Gamma}{V})\frac{d}{dt}=(3H+\frac{\Gamma}{V})\dot{\phi}\frac{d}{d\phi}=-\frac{V'}{V}\frac{d}{d\phi}
\end{eqnarray}

From above equation, Eq.(\ref{22}) and Eq.(\ref{25}), the expression $\frac{(\delta\phi)'}{\delta\phi}$ is obtained
\begin{eqnarray}\label{27}
\frac{(\delta\phi)'}{\delta\phi}=\frac{1}{(\ln V)'}[(\ln V)''+\dot{\phi}(\frac{\Gamma}{V})'+\frac{1}{2}(-\dot{\phi}\frac{\Gamma}{V}+2(\ln V)')\\
\nonumber
\times(\frac{V\dot{\phi}}{H})[1+\frac{\Gamma}{4HV}+\frac{\Gamma'\dot{\phi}}{48H^2V}](1-\frac{2V}{\rho_c})]~~~~~~~~~~~~~~
\end{eqnarray}

We will return to the above relation soon. Following Refs.\cite{1-m},  \cite{10-f}, \cite{6-f},  we introduce auxiliary function $\chi$ as
\begin{equation}\label{28}
\chi=\frac{\delta\phi}{(\ln V)'}\exp[\int\frac{1}{3H+\frac{\Gamma}{V}}(\frac{\Gamma}{V})'d\phi]
\end{equation}

From above definition we have
\begin{eqnarray}\label{29}
\frac{\chi'}{\chi}=\frac{(\delta\phi)'}{\delta\phi}-\frac{(\ln V)''}{(\ln V)'}+\frac{(\frac{\Gamma}{V})'}{3H+\frac{\Gamma}{V}}
\end{eqnarray}

Using above equation and Eq.(\ref{27}) we find
\begin{equation}\label{30}
\frac{\chi'}{\chi}=\frac{1}{2}(-\frac{\dot{\phi}}{(\ln V)'}\frac{\Gamma}{V}+2)
(\frac{V\dot{\phi}}{H})[1+\frac{\Gamma}{4HV}+\frac{\Gamma'\dot{\phi}}{48H^2V}](1-\frac{2V}{\rho_c})
\end{equation}
We could rewrite this equation, using Eqs. (\ref{7}) and (\ref{8})
\begin{equation}\label{31}
\frac{\chi'}{\chi}=-\frac{9}{8}\frac{2H+\frac{\Gamma}{V}}{(3H+\frac{\Gamma}{V})^2}(\Gamma+4HV-\frac{\Gamma'(\ln V)'}{12H(3H+\frac{\Gamma}{V})})\frac{(\ln V)'}{V}\frac{[1-\frac{2V}{\rho_c}]}{1-\frac{V}{\rho_c}}
\end{equation}

A solution for the above equation is
\begin{eqnarray}\label{32}
\chi(\phi)=C\exp(-\int\{-\frac{9}{8}\frac{2H+\frac{\Gamma}{V}}{(3H+\frac{\Gamma}{V})^2}~~~~~~~~~~~~~~~~~~\\
\nonumber \times(\Gamma+4HV-\frac{\Gamma'(\ln V)'}{12H(3H+\frac{\Gamma}{V})})\frac{(\ln V)'}{V}\frac{[1-\frac{2V}{\rho_c}]}{1-\frac{V}{\rho_c}}\}d\phi)
\end{eqnarray}

where $C$ is integration constant. From above equation and Eq.(\ref{29})  the change of variable $\delta\phi$ is determined
\begin{equation}\label{33}
\delta\phi=C(\ln V)'\exp(\Im(\phi))
\end{equation}

where
\begin{eqnarray}\label{34}
\Im(\phi)=-\int[\frac{(\frac{\Gamma}{V})'}{3H+\frac{\Gamma}{V}}+(\frac{9}{8}\frac{2H+\frac{\Gamma}{V}}{(3H+\frac{\Gamma}{V})^2}~~~~~~~~~~~~~~~~~~~~~~\\
\nonumber
\times(\Gamma+4HV-\frac{\Gamma'(\ln V)'}{12H(3H+\frac{\Gamma}{V})})\frac{(\ln V)'}{V}\frac{[1-\frac{2V}{\rho_c}]}{1-\frac{V}{\rho_c}})]d\phi
\end{eqnarray}
In the above calculations we have used the perturbation method in the warm inflation models \cite{1-m}, \cite{6-f}, \cite{10-f}, where the small change of variable $\delta\phi$ could be generated by thermal fluctuations instead of quantum fluctuations \cite{5},   and the integration constant $C$ may be driven by boundary conditions for field perturbation. Perturbed matter fields of our model are radiation $\delta\rho_r$, inflaton $\delta\phi$  and velocity $k^{-1}(P+\rho)v_{,i}$. We can explain the cosmological perturbations in terms of gauge-invariant variables. These variables are important for development of perturbation after the end of inflation period. The curvature perturbation $\mathfrak{R}$ and entropy perturbation $e$ are defied by \cite{nn-2}
\begin{eqnarray}\label{}
\mathfrak{R}=\Phi-k^{-1}aHv~~~~~~~\\
\nonumber
e=\delta P-c_s^2\delta\rho~~~~~~
\end{eqnarray}
where $c_s^2=\frac{\dot{P}}{\dot{\rho}}$. The boundary condition of warm inflation models are found in very large scale limits i.e., $k\ll aH$ where the curvature perturbation $\mathfrak{R}\sim const$ and the entropy perturbation vanishes \cite{nn-3}.

Finally the density perturbation is presented by \cite{12-f}
\begin{equation}\label{35}
\delta_H=\frac{16\pi}{5}\frac{\exp(-\Im(\phi))}{(\ln V)'}\delta\phi=\frac{16\pi}{15}\frac{\exp(-\Im(\phi))}{Hr\dot{\phi}}\delta\phi
\end{equation}

 In warm inflation model the fluctuations of the scalar field in high dissipative regime ($r\gg 1$) may be generated by thermal fluctuation instead of quantum fluctuations \cite{5} as
\begin{equation}\label{36}
(\delta\phi)^2\simeq\frac{k_F T_r}{2\pi^2}
\end{equation}

where in this limit freeze-out wave number $k_F=\sqrt{\frac{\Gamma H}{V}}=H\sqrt{3r}\geq H$ corresponds to the freeze-out scale at the point when, dissipation damps out to thermally excited fluctuations ($\frac{V''}{V'}<\frac{\Gamma H}{V}$) \cite{5}. With the help of the above equation and Eq.(\ref{35})  high dissipative regime ($r\gg 1$) we find
\begin{equation}\label{37}
\delta_H^2=\frac{128\sqrt{3}}{75}\frac{\exp(-2\tilde{\Im}(\phi))}{\sqrt{r}\tilde{\epsilon}}\frac{T_r}{H}
\end{equation}

where
\begin{equation}\label{38}
\tilde{\Im}(\phi)=-\int[\frac{1}{3Hr}(\frac{\Gamma}{V})'+\frac{9}{4}(1-\frac{(\ln\Gamma)'(\ln V)'}{36rH^2})(\ln V)'\frac{[1-\frac{2V}{\rho_c}]}{1-\frac{V}{\rho_c}}]d\phi
\end{equation}

and
\begin{equation}\label{39}
\tilde{\epsilon}=\frac{V'^2}{2rV^3}\frac{1-\frac{2V}{\rho_c}}{1-\frac{V}{\rho_c}}
\end{equation}

An important perturbation parameter is scalar index $n_s$ which
in high dissipative regime is given by
\begin{equation}\label{40}
n_s=1+\frac{d\ln \delta_H^2}{d\ln k}\approx
1-\frac{3}{4}\tilde{\epsilon}+\frac{3}{4}\tilde{\eta}+(\frac{\tilde{\epsilon}}{3r})^{\frac{1}{2}}(2\tilde{\Im}'(\phi)+\frac{r'}{2r})
\end{equation}

where
\begin{equation}\label{41}
\tilde{\eta}=\frac{2V'}{V^2r[1-\frac{V}{\rho_c}]}
[\frac{V''}{V'}-\frac{V'}{V}-\frac{r'}{2r}-\frac{V'}{\rho_c-2V}+\frac{V'}{2\rho_c-2V}]
\end{equation}
In Eq.(\ref{40}) we have used a relation between small change of
the number of e-folds and interval in wave number ($dN=-d\ln k$).
During inflation epoch,
there are two independent components of gravitational waves
($h_{\times +}$) with action of massless scalar field that are
produced by the generation of tensor perturbations. The amplitude
of tensor perturbation is given by
\begin{eqnarray}\label{43}
A_g^2=\frac{1}{4\pi}(\frac{H}{2\pi})^2\coth[\frac{k}{2T}]
\end{eqnarray}

where, the temperature $T$ in extra factor $\coth[\frac{k}{2T}]$,
denotes the temperature of the thermal background of
gravitational wave \cite{7}. Spectral index $n_g$ may be found as
\begin{eqnarray}\label{44}
n_g=\frac{d}{d\ln k}(\ln [\frac{A_g^2}{\coth(\frac{k}{2T})}])\simeq-2\tilde{\epsilon}
\end{eqnarray}
where $A_g\propto k^{n_g}\coth[\frac{k}{2T}]$ \cite{7}.  Using Eqs. (\ref{37})  and
(\ref{43}) we write the tensor-scalar ratio in high dissipative
regime
\begin{eqnarray}\label{45}
R(k)=\frac{A_g^2}{P_R}|_{k=k_{0}}
\end{eqnarray}

where $k_{0}$ is referred  to pivot point \cite{7} and $P_R=\frac{25}{4}\delta_H^2$. An upper bound for this parameter is obtained
by using    WMAP9 and BICEP2 observational data, $R<0.36$ \cite{6}.

\section{Exponential potential }
In this section we consider our model with the tachyonic
effective potential
\begin{equation}\label{27}
V(\phi)=V_0\exp(-\alpha\phi)
\end{equation}

where parameter $\alpha>0$ (with unit $m_p$) is related to mass
of tachyon field \cite{8}. The exponential form of potential have
characteristics of tachyon field ($\frac{dV}{d\phi}<0$ and
$V(\phi\rightarrow 0)\rightarrow V_{max}$ ). We develop our model
in high dissipative regime i.e. $r\gg 1$ for a constant
dissipation coefficient $\Gamma$. By using Eq.(\ref{8}) and
potential (\ref{27}), the scalar field in terms cosmic time is
found
\begin{equation}\label{28}
\phi(t)=\frac{1}{\alpha}\ln[\exp(\alpha\phi_i)+\frac{\alpha^2V_0}{\Gamma}t]
\end{equation}

where $\phi(t=t_i=0)=\phi_i$. Using above equation, Eqs.(\ref{7})
and (\ref{27}) we find the potential and Hubble parameter as
\begin{eqnarray}\label{29}
V(t)=\frac{V_0}{\exp(\alpha\phi_i)+\frac{\alpha^2V_0}{\Gamma}t}~~~~~~~~~~~~~~~~~~~~~~~~~~\\
\nonumber
H^2=\frac{1}{3}\frac{V_0}{\exp(\alpha_i)+\frac{\alpha^2V_0}{\Gamma}t}[1-\frac{V_0/\rho_c}{\exp(\alpha\phi_i)+\frac{\alpha^2V_0}{\Gamma_0}t}]
\end{eqnarray}

Dissipative parameter $r=\frac{\Gamma}{3HV}$ in this case becomes
\begin{equation}\label{30}
r=\frac{\Gamma_0}{\sqrt{3}}\frac{(\exp(\alpha\phi_i)
+\alpha^2\frac{V_0}{\Gamma_0}t)^{\frac{3}{2}}}{V_0^{\frac{3}{2}}(1-\frac{\exp(\alpha\phi_i)
+\alpha^2\frac{V_0}{\Gamma_0}t}{V_0\rho_c})^{\frac{1}{2}}}
\end{equation}

Using Eq.(\ref{9}) we find a relation between the energy densities
of radiation and inflaton fields.
\begin{equation}\label{31}
\rho_{\gamma}=\frac{3\rho_{\phi}^{\frac{3}{2}}}{4\sqrt{3}\Gamma_0(1-\frac{\rho_{\phi}}{\rho_c})^{\frac{1}{2}}}
\end{equation}

Power-spectrum in this case becomes (from Eq.(\ref{19}))
\begin{equation}\label{32}
P_R=\frac{3^{\frac{5}{4}}\exp(-2\tilde{\Im}(\phi))\Gamma_0^{\frac{5}{4}}T_r}{2\pi^2\alpha^2}V^{\frac{5}{4}}(1-\nu)^{\frac{5}{4}}
\end{equation}

where $\nu=\frac{V}{\rho_c}$ describes the quantum geometry
effects in LQC and $\tilde{\Im}(\phi)=-\frac{9}{4}\ln V$. From Eq.(\ref{26}) we find the tensor-scalar
ratio as
\begin{equation}\label{33}
R=\frac{4\exp(2\tilde{\Im}(\phi))\times
3^{\frac{1}{4}}\alpha^2}{3T_r\Gamma_0^{\frac{5}{2}}}\frac{V^{\frac{7}{4}}}{(1-\nu)^{\frac{1}{4}}}\coth[\frac{k}{2T}]
\end{equation}

From  observational data, we know $P_R=2.28\times 10^{-9}$ and
$R=0.21<0.36$ \cite{6}. From above equations and WMAP7 data we
find an upper bound for the potential
\begin{equation}\label{34}
V_{*}<3.4\times 10^{-4}
\end{equation}

We have obtained above equation in $\nu<1$ limit.By using BICEP2 data, we have found a new maximum of $V_{*}$ (See for example \cite{end}).

%\begin{figure}
%\includegraphics{Vahid07.eps
%}
%\caption{We plot the parameter $m^2$ in term of dissipation parameter $\Gamma_0$ where $T=T_r=2.24\times 10^{16} GeV, K_{*}=0.002 Mpc^{-1} $ and $\kappa=1$  %}\label{fk3}
%\end{figure}
\section{Conclusion}
Tachyon  inflation model  with  overlasting form of potential
$V(\phi)=V_0\exp(-\alpha\phi)$ which agrees with tachyon
potential properties have been studied. The main problem of
inflation theory is how to attach the universe to the end of the
inflation period. One of the solutions of this problem is the
study of inflation in the context of warm inflation \cite{3}. In
this model radiation is produced during inflation period where its
energy density is kept nearly constant. This is phenomenologically
fulfilled by introducing the dissipation term $\Gamma$. The study
of warm inflation model as a mechanism that gives an end for
tachyon inflation are motivated us to consider the warm tachyon
inflation model. In this article we have considered warm-tachyon
inflationary universe  model in the framework of effective field
theory LQC. In slow-roll approximation the explicit expressions
for the tensor-scalar ratio $R$, scalar spectrum $P_R$ and  index
$n_s$  have been presented. We have
developed our specific model by exponential potential. In this
case we have presented perturbation parameters and constrained this
parameters by  observational data. We also have constrained
the exponential potential  by using these data.

\end{document}